\RequirePackage{lineno}
\documentclass[aps,prl,superscriptaddress,twocolumn]{revtex4-1}

\usepackage{graphicx,amsmath}
\usepackage{color}

\usepackage{booktabs}
\usepackage[flushleft]{threeparttable}

\usepackage{amssymb}
\usepackage{amsmath}
\usepackage{float}
\usepackage{psfrag}
\usepackage{siunitx}
\usepackage{ulem}

\usepackage{natbib}
\def\be{\begin{equation}}
\def\ee{\end{equation}}
\def\ba{\begin{eqnarray}}
\def\ea{\end{eqnarray}}

\begin{document}
%\linenumbers

\title{Divergence of critical fluctuations on approaching catastrophic phase inversion in turbulent emulsions}

\author{Lei Yi}
\affiliation{Center for Combustion Energy, Key Laboratory for Thermal Science and Power Engineering of MoE, Department of Energy and Power Engineering, Tsinghua University, 100084 Beijing, China}

\author{Ivan Girotto}
\affiliation{Department of Physics and Department of Mathematics and Computer Science and J. M.
	Burgers Centre for Fluid Dynamics, Eindhoven University of Technology, 5600 MB
	Eindhoven, The Netherlands}
\affiliation{International Center for Theoretical Physics, 34151, Trieste, Italy}

\author{Federico Toschi}
\thanks{Email: f.toschi@tue.nl}
\affiliation{Department of Physics and Department of Mathematics and Computer Science and J. M.
	Burgers Centre for Fluid Dynamics, Eindhoven University of Technology, 5600 MB
	Eindhoven, The Netherlands}
\affiliation{CNR-IAC, Via dei Taurini 19, 00185 Rome, Italy}

\author{Chao Sun}
\thanks{Email: chaosun@tsinghua.edu.cn}
\affiliation{Center for Combustion Energy, Key Laboratory for Thermal Science and Power Engineering of MoE, Department of Energy and Power Engineering, Tsinghua University, 100084 Beijing, China}
\affiliation{Department of Engineering Mechanics, School of Aerospace Engineering, Tsinghua University, Beijing 100084, China}

\date{\today}

\begin{abstract} 

{
Catastrophic phase inversion, the sudden breakdown of a dense emulsion, occurs when the dispersed majority phase irreversibly exchanges role with the continuous minority phase. This common process has been extensively studied over the past decades and yet its fundamental physical mechanism has remained largely unexplored.
Here we experimentally and numerically study the dynamics of catastrophic phase inversion as it occurs when the volume fraction of the dispersed phase exceeds a critical volume fraction (typically around $92\%$ in experiments). Our data accurately quantify the abrupt change of both the global torque and average droplet size at approaching and across the phase inversion point, exhibiting strong hysteresis.
Most importantly, we reveal that the fluctuations in the global torque diverge as a power-law while approaching the critical volume-fraction and we connect their growth to the formation of highly heterogeneous spatial droplet structures. 
The present finding, unveiling the tight connection between fluctuations in dynamic heterogeneity and the critical divergence of torque fluctuation, paves the way to a quantitative description of catastrophic phase inversion as an out-of-equilibrium critical-like phenomena.
}

\end{abstract}

%\pacs{47.55.D-, 66.10.C-}

\maketitle

%\annotation{MANUSCRIPT FORMATTING GUIDE: Section 1: beginning with up to 500 words of referenced text expanding on the background to the work (some overlap with the summary is acceptable)}

\section{Introduction} 

Emulsions, metastable mixtures of two immiscible liquids,
are ubiquitous both in ordinary household products (e.g., mayonnaise) as well
as in the chemical, pharmaceutical, food, and cosmetics
industries~\cite{leal2007emulsion}.
Catastrophic phase inversion in emulsions describes the phenomenon whereby the high-volume-fraction dispersed droplet phase rapidly converts into the continuous phase, and vice versa. While phase inversion can be triggered in many ways, including temperature or chemical changes, here we limit to catastrophic phase inversion triggered by a volumetric change in the relative fraction of the emulsion oil-to-water ratio~\cite{perazzo2015phase}.
While catastrophic phase inversion is easy to identify, due to the abrupt changes in the morphology, rheology, and stability of the system~\cite{yeo2000phase}, its underlying dynamics has not been fully elucidated. Due to the rapidity of the process, the dynamics of catastrophic phase inversion precursors have been difficult to be identified and studied.

It has been widely observed that the phase inversion is preceded by the apparent divergence of the effective viscosity, which rapidly and dramatically drops with the formation of the new continuous phase~\cite{tyrode2005emulsion,ioannou2005phase,sousa2021properties,pouplin2011wall,bakhuis2021catastrophic}.
Experiments in an oil-water pipe flow have found that the phase inversion was accompanied by large fluctuations in pressure gradient and in mixture impedance~\cite{ioannou2005phase,piela2006experimental}, however, until now, these observations lack clear explanations and quantitative description.
Apart from the dramatic change in the macroscopic dynamics, i.e., in the emulsion rheology, abrupt transitions in the local morphological structures are found during the phase inversion through direct
visualization~\cite{yeo2000phase,piela2006experimental,piela2008phase} or via Laser-induced fluorescence
(LIF)~\cite{liu2005experimental,liu2006laser}.  
Morphological transitions during phase inversion share similar behaviors~\cite{piela2009phase}, e.g., the formation of giant anomalous droplets or multiple emulsions and development of the new continuous phase, whereas a unifying theoretical description of the change in phase behaviors remains unexplored.  
One of the main reasons for this lack of understanding is the limited number of studies able to quantitatively analyze the evolution of the dispersed droplets statistics, due to the challenges associated with dynamically characterizing flowing emulsions in the high volume-fraction regimes at which the phase inversion typically occurs~\cite{kumar2015recent}.

Many models have been developed to describe catastrophic phase inversion in the past decades.
The widely-used catastrophe theory contributed the term “catastrophic” in the phase inversion, and while it was found to successfully describe qualitative features, it could not provide satisfactory quantitative predictions~\citep{dickinson1981interpretation,vaessen1995applicability}.
Another mechanism to predict the phase inversion is based on the criterion of minimal energy in the system, which requires characterizing the droplet dynamics in the pre- and post-phase-inversion emulsion and is therefore challenging~\cite{brauner2002modeling,yeo2002simple}.
Many investigations have been performed to study droplet dynamics across the phase inversion in emulsions, yielding models based on the competition between droplet breakup and coalescence~\cite{nienow2004break,poesio2008minimal}.
Recently, an extended Ginzburg-Landau model was proposed to describe phase inversion in pipe flows, in analogy with the classical mean-field theory of phase transitions in thermodynamics~\cite{piela2009phenomenological}. 
Notwithstanding extensive efforts, satisfactory models are generally not available and the physical picture of phase inversion tightly connecting global and local dynamics remains largely unexplored and unclear.
One of the key reasons is the high number of variables involved in the phase inversion process, including the concentration of oil and the water phases, flow type and intensity, vessel geometry, temperature, and
presence of surfactant, among others~\citep{perazzo2015phase}.

\textcolor{black}{
Here we investigate the dynamics of catastrophic phase inversion in a turbulent emulsion by varying the oil-to-water ratio, focusing on the evolutions of both the macroscopic rheological properties and microscopic droplet statistics across the phase inversion.
We quantitatively show that catastrophic phase inversion is characterized by the divergences of both the fluctuations in the global torque and the variations in the local droplet statistics at approaching the critical oil volume fraction at which the phase inversion occurs.  Our study provides an alternative approach to model the phase inversion as a critical-like phenomena.
}

\section*{Results and Discussion}

\subsection*{Global torque measurements}

\begin{figure*}[!htb]
	\centering
	\includegraphics[width = 1\textwidth]{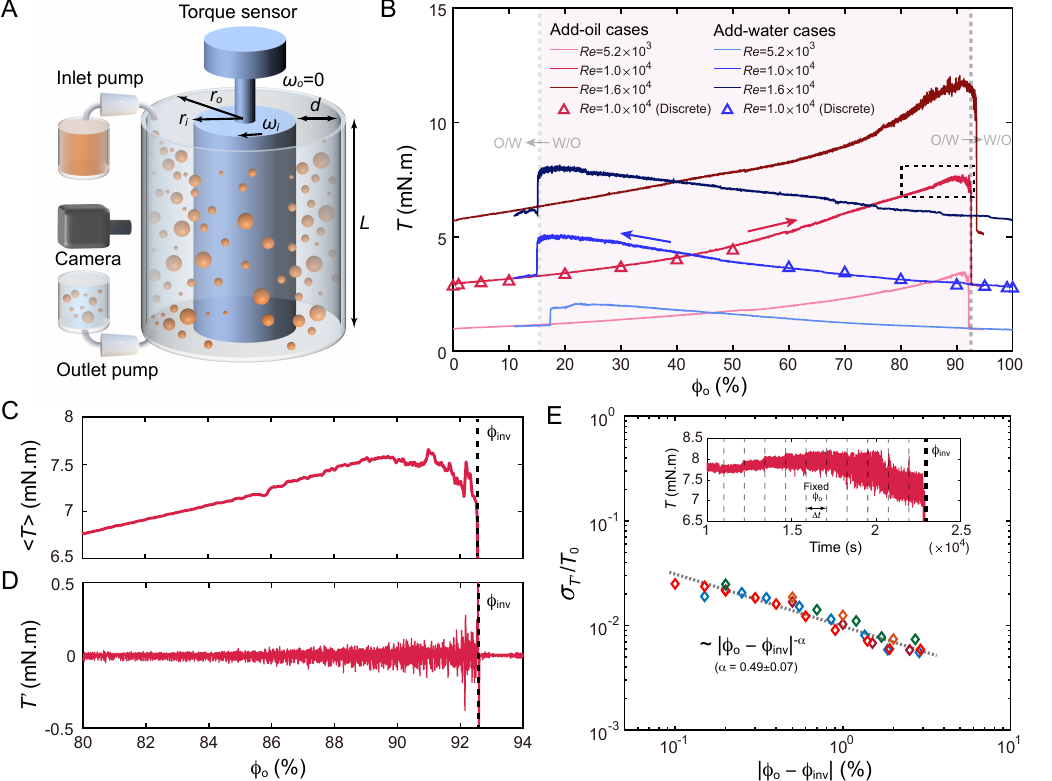}
	\caption{
	(A) Sketch of the experimental set-up. The emulsion was generated in the gap by rotating the inner cylinder at a constant angular velocity $\omega_{i}$.
	The torque exerted on the inner cylinder was recorded by the torque sensor. 
	The high-speed camera was used to capture the morphological structures of the emulsion.
	Two micropumps were employed to gradually increase the volume fraction of the dispersed phase in the emulsion (SI Appendix).
	(B) The torque $T$ versus the oil volume fraction, $\phi_{o}$, at various Reynolds numbers. The lines denote the continuous experiments (red lines for add-oil cases and blue for add-water cases), while the upper triangles represent the discrete experiments (red for O/W emulsions and blue for W/O).
	The shaded area between the two dashed vertical lines denotes the hysteresis region, bounded by the catastrophic phase inversions: from W/O to O/W (left direction) and from O/W to W/O (right direction).
	(C) The mean component of the torque $\left<T\right>$ versus the oil volume fraction $\phi_{o}$. 
	The vertical black dashed line gives the volume fraction at which the phase inversion occurs ($\phi_{inv}$).
	Note that the data in C and D correspond to the dashed black box region in panel B for the add-oil case at $Re = 1.0\times10^{4}$. 
	(D) The torque fluctuation $T^{\prime}$ versus $\phi_{o}$. This panel shows the growing behavior of the torque fluctuation towards the phase inversion point $\phi_{inv}$.
\textcolor{black}{
	(E) The normalized standard deviation of the torque fluctuation $\sigma_{_{T^{\prime}}}/T_{0}$ as a function of $|\phi_{o} - \phi_{inv}|$ for the add-oil case at $Re = 1.0\times10^{4}$.
	The various colors represent datasets from $5$ repeated experiments.
	The standard deviations are normalized by the mean torque measured just before the phase inversion $T_{0}$.
	 The dashed line denotes a power-law dependence: $\sigma_{_{T^{\prime}}}/T_{0} \sim |\phi_{o} - \phi_{inv}|^{-\alpha}$, and the fitted exponent is found to be $\alpha = 0.49\pm 0.07$.
	The inset shows a typical torque-time series to measure the torque fluctuations at approaching the phase inversion.
	Step-by-step measurements of the torque are performed with the duration $\Delta t = \SI{20}{mins}$ per measurement step at a fixed $\phi_{o}$ and by gradually increasing $\phi_{o}$, step by step, until the phase inversion occurs.
	}
	}
	\label{fig1}
\end{figure*}

We start by characterizing the rheological properties of the flowing oil-water emulsion using time-resolved measurements of the global torque, $T$, necessary to maintain a constant rotating angular velocity, $\omega_{i}$, for the inner cylinder of a Taylor-Couette (TC) system (see Fig.~\ref{fig1}$A$).
The experimental TC system is characterized by an inner cylinder of radius $r_{i} = 25\rm~mm$, an outer cylinder of radius $r_{o} = 35\rm~mm$, a gap $d = r_{o} - r_{i} = 10\rm~mm$ and a height $L = 75\rm~mm$.
The outer cylinder is stationary, while the inner cylinder rotates and stirs the mixture of two immiscible liquids (i.e. oil and water phases) in the gap to generate a turbulent emulsion. 
A control parameter of the system is the Reynolds number, defined as: 
\begin{equation}
	Re = \omega_{i}r_{i}d/\nu_{w},
\end{equation}
where $\nu_{w}$ is the kinematic viscosity of the water phase.
The second control parameter is the volume fraction of oil, given by: 
\begin{equation}
\phi_{o} = V_{o} / (V_{o} + V_{w}), 
\end{equation}
where $V_{o}$ and $V_{w}$ denote the volumes of the oil and water phases, respectively.

We report in Fig.~\ref{fig1}$B$ the global torque $T$ as a function of the oil volume fraction $\phi_{o}$.
We have performed two kinds of experiments: $1)$ continuous experiments, where we continuously, and slowly vary the oil volume fraction $\phi_{o}$, and $2)$ discrete experiments, in which oil and water phases are added to the system at a fixed $\phi_{o}$ before starting with the stirring (Materials and Methods, SI Appendix).
First, we show the results of discrete experiments at $Re = 1.0\times10^{4}$ (see triangles in Fig.~\ref{fig1}$B$). 
We find that, as the oil volume fraction is less than $50\%$ ($\phi_{o} \leqslant 50\%$), the water phase always dominates the system and, consequently, acts as the continuous phase, giving the oil-in-water (O/W) emulsions. The required torque increases with $\phi_{o}$ (see red triangles in Fig.~\ref{fig1}$B$), suggesting an increase in the effective viscosity of the emulsion.
For $\phi_{o} \geqslant 60\%$, the oil phase switches to the continuous phase in the system and thus we observe W/O emulsions. The increasing trend of the torque can also be measured along the opposite direction, i.e., at increasing the volume fraction of the dispersed water phase (see blue triangles in Fig.~\ref{fig1}$B$). 

Next, we focus on the continuous experiments at the same Reynolds number, i.e., $Re = 1.0\times10^{4}$. 
According to the type of liquid added into the system to vary the volume fraction $\phi_{o}$, the continuous experiments can be divided into two groups: $1)$ the add-oil (starting from $\phi_{o} = 0\%$) and $2)$ add-water (starting from $\phi_{o} = 100\%$) cases, respectively (more details in SI Appendix).
 As shown by the red line (accompanied by an arrow) in the middle position of Fig.~\ref{fig1}$B$, the torque for the O/W emulsion increases at increasing $\phi_{o}$ (add-oil case), and the torque values are consistent with those in the discrete experiments for $\phi_{o} \leqslant 50\%$. This increasing trend continues until $\phi_{o} \approx 92\%$, where the torque undergoes a sharp drop of over $60\%$, clearly indicating the macroscopic signature of the occurrence of catastrophic phase inversion from O/W to W/O emulsion.
 Here, we define the oil volume fraction in which the catastrophic phase inversion occurs as the phase inversion point, $\phi_{inv}$.
 We stress that, in the continuous experiments, in contrast to the discrete experiments, the phase inversion point can be postponed up to a high value, around $92\%$, by gradually (quasi-statically) increasing the dispersed-phase volume fraction.
 After the phase inversion has occurred, the torque value settles to that of the corresponding water-in-oil (W/O) emulsion.
 
Regarding the add-water continuous experiment (blue line accompanied by an arrow in Fig.~\ref{fig1}$B$), we find a similarly increasing trend for the torque but in the opposite direction. However, in this case, the torque suddenly jumps down around $\phi_{o} \approx 17\%$, a very different phase inversion point from the add-oil case, suggesting that phase inversion exhibits strong hysteresis.
Thus, the system can exist as either O/W or W/O states for a wide oil volume-fraction range of around $80\%$ ($\phi_{o} \approx 17\%$ to $\phi_{o} \approx 92\%$, shaded area in Fig.~\ref{fig1}$B$).
The wide hysteresis region implies that we can set the dispersed/continuous phase at a given volume fraction before the phase inversion by choosing the initial dispersed phase.
Furthermore, for various Reynolds numbers, the global torques show similarly growing behaviors followed by a sharp decrease at the phase inversion point.
It is also found that the critical phase inversion point, $\phi_{inv}$, hardly changes with $Re$ in the current parameter regime, for both add-oil and add-water cases.
This result deserves further investigation as we know that the local droplet size decreases at increasing the Reynolds number.
Next, we focus on the torque signal near to the phase inversion point to gain insights into the nature of the dynamics through which the global torque sharply drops by more than $60\%$.

\subsection*{Divergence of critical torque fluctuations towards the phase inversion}
By looking at the torque signal in the vicinity of the phase inversion (black dashed box in Fig.~\ref{fig1}$B$), we observe that the torque exhibits a slight reduction just before the phase inversion.
Moreover, we clearly see that the torque is accompanied by large fluctuations on approaching the phase inversion point.
The torque can be decomposed as $T=\langle T\rangle+T^{\prime}$, where $\left<T\right>$ is the mean component (time-averaged) while $T^{\prime}$ the fluctuation component (Section B, SI Appendix).
Taking the add-oil case at $Re = 1.0\times10^{4}$ as an example, we plot the mean torque $\left<T\right>$ versus $\phi_{o}$ in Fig.~\ref{fig1}$C$, which clearly shows that the mean torque first increases, then undergoes a slight reduction, and finally drops down at $\phi_{inv}$.

Fig.~\ref{fig1}$D$ gives the fluctuation component of the torque, $T^{\prime}$, as a function of $\phi_{o}$, which shows how the fluctuations become progressively larger approaching the phase inversion point, but reduce significantly to a very small value after the phase inversion.
We now focus on these torque fluctuations to investigate the origin of phase inversion.
\textcolor{black}{
To quantitatively characterize how the torque fluctuation changes near to the phase inversion, we perform step-by-step measurements of the torque with a duration of $\Delta t = \SI{20}{mins}$ per measurement step at fixed $\phi_{o}$, and then by gradually increasing $\phi_{o}$ step by step until the phase inversion occurs (see the inset in Fig.~\ref{fig1}$E$).
}
The torque fluctuation monotonically increases as $\phi_{inv}$ is approached, in a manner similar to critical phenomena, in which the dynamics of the system is typically governed by the power-law scaling~\citep{muller2015marginal,sastry2016critically,charbonneau2017glass}.
Motivated by this, we define the volume-fraction distance to the critical phase inversion point as $|\phi_{o} - \phi_{inv}|$, and plot the normalized standard deviation of the torque fluctuation $\sigma_{_{T^{\prime}}}/T_{0}$ ($T_{0}$ is the mean torque value just before the phase inversion) versus $|\phi_{o} - \phi_{inv}|$ (see Fig.~\ref{fig1}$E$ main panel).
We find that, $\sigma_{_{T^{\prime}}}/T_{0}$ shows a power-law dependence, i.e.,
 \begin{equation}
	\sigma_{_{T^{\prime}}}/T_{0} \sim |\phi_{o} - \phi_{inv}|^{-\alpha},
\end{equation}
where $\alpha$ is found to be $\alpha = 0.49\pm 0.07$, suggesting that the fluctuations of global torque diverge as $\phi_{inv}$ is approached.
\textcolor{black}{
Therefore, we show that the sudden drop of the torque follows the critical divergence of its fluctuations.
}

\subsection*{Local droplet statistics} 
\begin{figure*}[!htb]
	\centering
	\includegraphics[width = 1\textwidth]{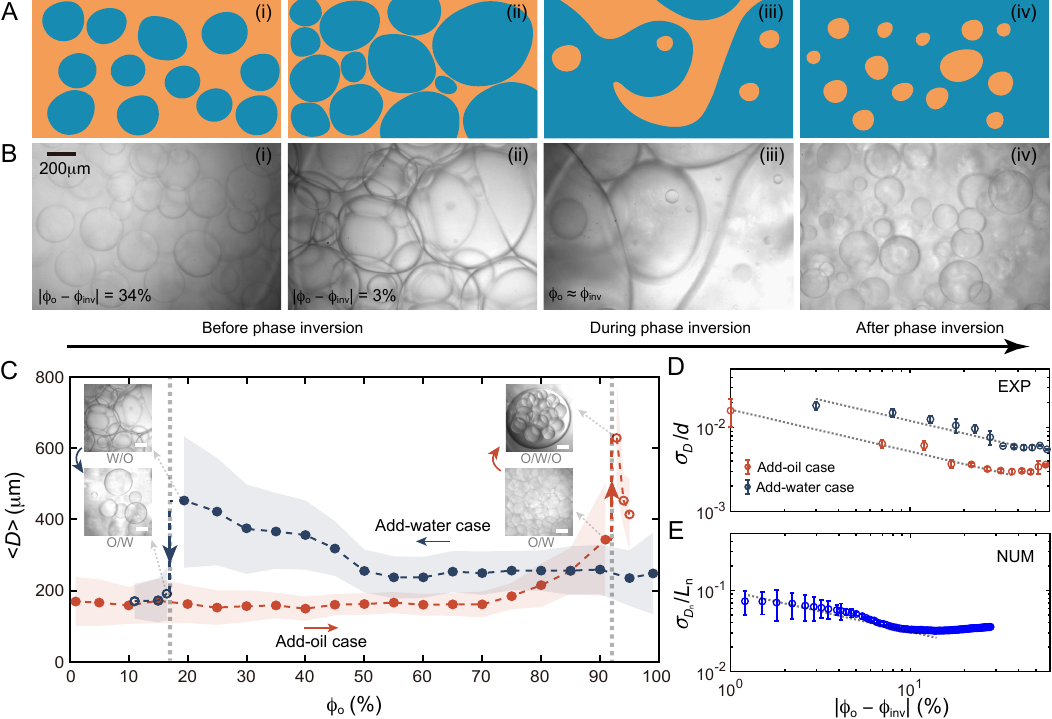}
 \vspace{-2mm}
	\caption{
	Evolutions of local droplet morphological structures and statistics.
	(A) A sketch to illustrate the phase inversion process.
	(B) The experimental snapshots for the phase inversion.
	From (i) to (iii), some giant droplets (or continuous regions) form due to the fast coalescence of neighboring droplets, indicating the development of heterogeneous droplet structures in the system.
	The scalar bar $\SI{200}{\mu m}$ is the same for (i-iv).
	(C) The average droplet diameter $\left<D\right>$ as a function of the oil volume fraction for both the add-oil case (red line) and add-water case (blue line) at $Re = 5.2\times10^{3}$. 
	The colored bands span one standard deviation around the average diameter.
	The vertical dashed lines with arrows denote the occurrence of the catastrophic phase inversion.
	The snapshots of the emulsion before and after the phase inversion are shown as insets. All scalar bars in these four inset snapshots represent $\SI{200}{\mu m}$.
    (D) The standard deviation of the droplet diameter normalized by the gap width $\sigma_{_D}/d$ as a function of the volume-fraction distance towards the critical phase inversion point $|\phi_{o} - \phi_{inv}|$ (Experimental results, EXP). 
    Dashed lines in panels D and E denote the power-law scaling $\sigma_{_D}/d \sim |\phi_{o} - \phi_{inv}|^{-\beta}$, where the fitted exponents $\beta$ for add-oil and add-water cases are $0.48\pm0.08$ and $0.46\pm0.1$, respectively.
	(E) The standard deviation of the droplet diameter normalized by the domain size in the simulations $\sigma_{_{D_{n}}}/L_{n}$ versus $|\phi_{o} - \phi_{inv}|$ (Numerical results, NUM).
	The dashed line resprents the scaling, $\sim |\phi_{o} - \phi_{inv}|^{-\beta}$, where  $\beta = 0.38 \pm 0.05$ by fitting.
	}
	\label{fig2}
\end{figure*}

To gain a better understanding of the global dynamics at approaching the phase inversion, we investigate the evolution of the local droplet structures before, during, and after the phase inversion (see Fig.~\ref{fig2}).
We present both the morphological transition of dispersed/continuous phase in Fig.~\ref{fig2}$B$, as well as the quantitative evolution of the sizes of the dispersed droplets in Fig.~\ref{fig2}$C$.
For both the add-oil and the add-water cases, the average droplet diameter $\left<D\right>$ is almost independent of the oil volume fraction, $\phi_{o}$, for relatively low volume-fraction regimes, away from the phase inversion point (see the typical snapshot in (Fig.~\ref{fig2}$B, \romannumeral1$)).
Here, the droplet diameter is the area equivalent diameter determined by the droplet area from imaging.
As $\phi_{inv}$ is approached, the droplet size increases, which can be associated to the dominant role that coalescence plays in the droplet dynamics in these high-volume-fraction regimes.
Note that only simple droplets are present before phase inversion and no double emulsion is observed, i.e., oil-in-water-in-oil (O/W/O) or water-in-oil-in-water (W/O/W).
In a highly-concentrated granular system, the random close packing (RCP) state is an important reference, around which the system usually undergoes a jamming transition~\citep{charbonneau2017glass}.
\textcolor{black}{As the random close packing of spheres is still controversial, here we use $\phi_{p}=66\%$ as reference volume-fraction value for RCP in an emulsion, based on previous investigation on jammed emulsions consisting of polydisperse droplets~\citep{clusel2009granocentric}.}
We observe that the starting point of dispersed-phase volume fraction for the increase of droplet size is around $60\% \sim 70\%$ (Fig.~\ref{fig2}$C$), which is comparable to the value of RCP, i.e. $\phi_{p} = 66\%$, suggesting that the volume fraction is significant in determining the dominant role between coalescence and breakup.
The closely-packed droplets in the dense emulsion (Fig.~\ref{fig2}$B, \romannumeral2$) demand some kind of disjoining pressure for not fully coalescing. A possible source can be the unavoidable surface-active impurities accumulated at the liquid-liquid interfaces in an experimental system~\citep{yi2022physical,soligo2020effect}, which stabilizes the emulsion by inhibiting the droplet coalescence, while the details of the stabilization mechanism deserve further investigation.
\begin{figure*}[!htb]
	\centering
	\includegraphics[width = 1\textwidth]{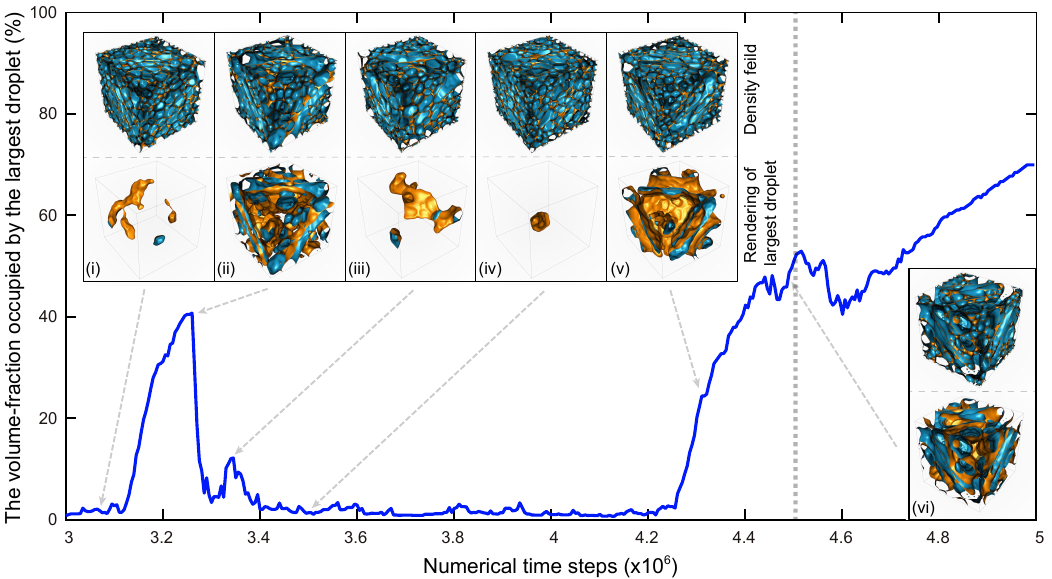}
	\caption{
	\textcolor{black}{
 A direct visualization of the tight connection between the droplet morphology and the presence of catastrophic phase inversion.
	The solid blue line represents the volume fraction occupied by the largest droplet in a very dense emulsion (with a constant droplet phase volume fraction around $77\%$) versus time in a fully resolved numerical simulation.
	 For the entire simulation time, the dense emulsion is forced with an almost critical large-scale forcing that can likely lead to phase inversion. 
	 The insets show density contour renderings from successive 3D numerical snapshots at different time, clearly displaying the evolution of the largest droplet present in the emulsion (lower-half) coupled with the density field for the entire numerical domain (upper-half).
	 The vertical dashed line denotes the position where the largest droplet occupies $50\%$ of the total volume fraction, indicating the start of the phase inversion.
	}
	}
	\label{fig3}
\end{figure*}

Moreover, at the edge of the phase inversion point, the emulsion becomes so dense that some giant droplets (see Fig.~\ref{fig2}$B, \romannumeral2$ and upper-half inset in Fig.~\ref{fig3}) are formed due to the merging of neighboring dispersed smaller droplets, giving rise to the enhanced spatial heterogeneity characterized by the increase of standard deviation of the droplet size (see the colored bands in Fig.~\ref{fig2}$C$).
These giant droplets of the dispersed phase grow by coalescing with other droplets and contribute to the formation of the continuous regions, which quickly develop and finally span the entire domain as the new continuous phase of the system (Fig.~\ref{fig2}$A, \romannumeral3$ and Fig.~\ref{fig2}$B, \romannumeral3$), thus resulting in the completion of the phase inversion (Section D in SI Appendix, more experimental snapshots).
\textcolor{black}{
The formation of giant droplets/regions of dispersed phase also explains the aforementioned observation that the mean torque exhibits a slight reduction very close to the phase inversion point in Fig.~\ref{fig1}$C$.
We roughly assume that the dispersed phase has a viscosity of bare fluid, while the rest of the domain is occupied by the dense emulsion with a much higher effective viscosity. 
Therefore, in the vicinity of the phase inversion, the effective viscosity of the entire system decreases owing to the growing of the continuous regions of the dispersed phase, thus resulting in the reduction of the global torque.
}

At the inversion, we observe the abrupt change of the average droplet size in Fig.~\ref{fig2}$C$, which directly corresponds to the jump of the global torque in Fig.~\ref{fig1}$B$.
Notably, we find that the aforementioned morphological transition occurs during a short time of around $\SI{5}{seconds}$, which is remarkable as a rough estimation of droplet number is on the order of $\mathcal{O}(10^{4})$, thus giving an average coalescence frequency of around $\SI{2}{kHz}$.
The mechanism of fast destabilization of emulsions in the phase inversion may be interpreted by the cascading propagation of the droplet coalescence inside the highly-concentrated emulsions under continuous shearing~\citep{bremond2011propagation,raj2017averaged,dekker2021creep}.
Note that double emulsion structures usually appear during the phase inversion process (see Fig.~S4 in SI Appendix). 
In the add-water case, those unstable double-emulsion structures quickly disappear after the phase inversion, under the action of turbulent flow, but, in the add-oil case, as the double emulsions were not completely destroyed during the limited time after phase inversion in experiments, the droplet size did not settle to the corresponding value for the W/O system (Fig.~\ref{fig2}$C$, O/W/O structures after the inversion).
\textcolor{black}{
We emphasize that, across the phase inversion, the dense emulsion is always in a dynamical equilibrium with droplet breakup and coalescence events balancing with each other.
To study the growth dynamics of spatial heterogeneity in the droplet structure, we performed numerical simulations of a turbulent emulsion, very close to the phase inversion, and maintained at a constant dispersed phase volume-fraction, quite similar to the long-time experimental measurement at fixed volume fraction (inset in Fig.~\ref{fig1}$E$).
The droplet with the largest volume is then chosen as an indicator of the local droplet structure. Fig.~\ref{fig3} shows the evolution of the volume-fraction occupied by the largest droplet in an emulsion as a function of time. 
The largest droplet first expands to occupy around $40\%$ of the entire domain volume due to coalescence, then it shrinks back to a lower value due to the fragmentation, and finally it grows again to dominate the system volume, resulting in the phase inversion.
This clearly illustrates the key role of the fluctuations in droplet structures near to the phase inversion.
}
A natural question arises as to whether this dynamic spatial heterogeneity of the droplet structure at the small scale correlates with the critical-like behaviors of torque fluctuation at the large scale on approaching the phase inversion.
We uncover this underlying connection in the next section.

\subsection*{The physical mechanism of phase inversion} 

\textcolor{black}{
Similar to the investigation of torque fluctuation, we expect that the dynamic spatial heterogeneity of droplet structure can be quantitatively characterized by the standard deviation of droplet size.
We display the standard deviation of droplet size rescaled by the gap width $\sigma_{_D}/d$ versus the distance to the critical phase inversion point $|\phi_{o} - \phi_{inv}|$.
As shown in Fig.~\ref{fig2}$D$, near the phase inversion point, it is found that $\sigma_{_D}/d$ increases monotonically on approaching the phase inversion, which can be roughly described by a power-law dependence, i.e., $\sigma_{_D}/d \sim |\phi_{o} - \phi_{inv}|^{-\beta}$, where the fitted exponents $\beta$ for add-oil and add-water cases are $0.48\pm0.08$ and $0.46\pm0.1$, respectively.
This suggests the diverging behavior of the spatial heterogeneity of droplet structure towards the critical phase inversion point.
Moreover, the critical diverging behavior of the droplet size on approaching the phase inversion is verified by 3-dimensional numerical simulations as well (Fig.~\ref{fig2}$E$), in which the droplet size is calculated based on the droplet volume.
}

\textcolor{black}{
Interestingly, by reinspecting Fig.~\ref{fig1}$E$, we find that both $\sigma_{_{T^{\prime}}}$ and $\sigma_{_D}$ diverge approaching the phase inversion and share similar power-law scalings as a function of $|\phi_{o} - \phi_{inv}|$, suggesting a correlation between the torque fluctuation and the dynamic change of droplet structures.
The dynamic behavior of droplet structures is shown by Fig.~\ref{fig3}, in which the largest droplet in a dense emulsion undergoes the fluctuation of its volume (lower-half inset (i-v) in Fig.~\ref{fig3}), at a constant dispersed phase volume fraction.
This dynamic variation of heterogeneous droplet structures is determined by the competition between the formation and fragmentation of the dispersed droplets/regions, and can be closely linked to the critical fluctuations of global torque (inset in Fig.~\ref{fig1}$E$).
Therefore, we suggest a description of catastrophic phase inversion as a critical-like phenomena characterized by both temporal fluctuations of global torque and dynamic spatial variations of heterogenous structures of the local droplet phase.
}

\textcolor{black}{
The aforementioned critical dynamics in the phase inversion, are largely reminiscent of the critical phenomenon in an equilibrium phase transition, in which an important characteristic is that the heterogeneous nature of dynamics is governed by power-law dependences on the distance to the critical point~\citep{huang2008statistical,domb1976phase,nishimori2011elements,keys2007measurement}.
It has been widely found that concentrated granular systems such as colloidal suspensions, foams, and emulsions, undergo the so-called jamming transition near the RCP, which has been extensively reported to share many features of the conventional phase transitions over the past decades~\citep{liu1998jamming, liu2001jamming, o2003jamming, lerner2012unified, denkov2009jamming, sastry2016critically, charbonneau2017glass}.
Here, owing to the turbulence present in the system, our study goes beyond the jamming regime to explore different packing configurations of an emulsion above RCP.
Motivated by this, we speculate that the critical dynamics near to the phase inversion point may be understood if the phase inversion is assumed to be analogous with the phase transition.
In a typical continuous phase-transition system (e.g., paramagnetic-ferromagnetic transition), the constituent particles are molecules with thermal noise, thus the key control parameter for the thermodynamics behaviors is temperature.
The critical phenomena associated with continuous phase transitions can be described by power laws with critical exponents, of which the typical one is the exponent describing the divergence of the thermal correlation length scale $\xi$, i.e., $\xi \sim |T - T_{c}|^{-\nu}$, where the exponent ${\nu = 1/2}$ is given by the Landau mean-field theory~\cite{landau2013statistical,domb1976phase,nishimori2011elements}.
These critical exponents in phase-transition systems can be linked to the power-law dependences we obtained for both the torque fluctuations and the droplet structure.
We note that as the thermal motion of the constituent droplets is too insignificant, dense emulsion systems are considered as athermal, and the role of noise is played by turbulent fluctuations.
The striking similarities of the critical dynamics between phase inversion systems and equilibrium phase transition should motivate theoretical works to establish a robust connection, and to help unravel the dynamics of the phase inversion via the framework of critical phenomena.
}

\section{Conclusions}
\vspace{-.3 cm}
In summary, we have experimentally and numerically investigated both the global rheological properties and local droplet statistics in a turbulent emulsion system undergoing catastrophic phase inversion.
We observe that the phase inversion is characterized by abrupt changes in both the global torque and local droplet size, which shows hysteresis behavior.
We reveal that the abrupt jump in the torque at the phase inversion point can be identified as a result of the divergence of torque fluctuations on approaching the critical volume fraction, and obeys a power-law relation.
The formation of giant and anomalous droplets/regions of dispersed phase results in spatially heterogeneous structures in the system before the phase inversion occurs.
We uncover that the growing dynamic variation of heterogeneous droplet structures at the small scale is tightly connected to the divergence of fluctuations in the global torque in the system.

The present research shows that the catastrophic phase inversion can be described as a critical-like phenomenon characterized by both temporal fluctuations of global torque and dynamic spatial variations of heterogeneity in droplet structures, providing a new approach to modeling the phase inversion.
The striking similarities between the phase inversion and the equilibrium phase transition regarding the dynamics in the vicinity of the critical point, should inspire further investigation on the underlying connections.
Moreover, our approach can be generalized to study the dynamics in a wide range of out-of-equilibrium soft-matter systems, such as foams and soft colloidal suspensions.

%\noindent {\bf Acknowledgements}
\section{Acknowledgements}
We thank Roberto Benzi, Stan van der Burg, Yibao Zhang, and Mingbo Li for their insightful discussions, and thank Cheng Wang for his help in the experiments.
This work is partially financially supported by the National Natural Science Foundation of China under grant no. 11988102 and the New Cornerstone Science Foundation through the XPLORER PRIZE.
Numerical simulations were performed thanks to granted PRACE projects (ID: 2018184340 \& 2019204899) along with CINECA and BSC for access to their HPC systems. 
This work was partially sponsored by NWO domain Science for the use of supercomputer facilities.

%\noindent{\bf
%	Author Information}
The authors declare no competing interests. 
%Readers are welcome to comment on the online version of the paper. 
%Correspondence and requests for materials should be addressed to.

%\subsection{Data availability} The data that support the findings of this study are available
%from the authors upon reasonable request.

\section{EXPERIMENTAL AND NUMERICAL METHODS}

Here, we provide basic information on experimental and numerical methods in this study.
For further details and additional figures, please refer to SI Appendix.

\subsection{1. Experiments}
%Appendix
%experiments
Two immiscible liquids were used in experiments, namely, silicone oil (density $\rho_{o}=866\rm~kg/m^{3}$, viscosity $\nu_{o}=2.1 \times10^{-6}\rm~m^{2}/s$) from Shin-Etsu and an aqueous ethanol-water mixture ($\rho_{w}=860\rm~kg/m^{3}$, $\nu_{w}=2.4\times10^{-6}\rm~m^{2}/s$) with $75\%$ ultra-pure water in volume from a water purification system (Milli-Q, Merck, Germany).
The densities of these two liquids are almost matched to eliminate the effect of the centrifugal force on the distribution of the dispersed droplets.
No surfactant is used in the experiments.
A circulating water bath is employed to keep the operating temperature at $22\pm0.1 ^{\circ}$C. The temperature gradient inside the emulsion is negligible due to the efficient mixing generated by the turbulent fluctuations~\citep{vangils2011rsi,grossmann2016high}.

Continuous experiments were performed by continuously, and very slowly, injecting the dispersed liquid into the system to vary the oil volume fraction $\phi_{o}$ using an inlet micropump (SI Appendix, Section A). At the same time, the well-mixed emulsion was pumped out of the system by an outlet micropump with the same flow rate of volume.
The global torque $T$ required to maintain the rotating inner cylinder at a given angular velocity $\omega_{i}$, is measured using a high-precision (up to $0.1\rm~nN\cdot m$) torque sensor (SI Appendix, Section B). 
The local droplet statistics of the emulsion were analyzed using high-speed videography.
Considering the meta-stability of the emulsion without adding surfactant, the emulsion coarsens and separates quickly as the external shear force is stopped. 
Thus, we performed direct visualization of the emulsion during the operation of the system, instead of taking the emulsion sample out for analysis.
A high-speed camera (Photron Fastcam NOVA S12) mounted with a long-distance microscope (Navitar) was used for characterizing the droplet structures (Fig.~\ref{fig2}$B$).
The droplet sizes are extracted from images using in-house Matlab codes and ImageJ software.

\subsection{2. Numerics}
%Numerics
In addition to the experiments, we also performed a series of computer simulations of the dense emulsions based on the Lattice Boltzmann method (LBM)~\citep{shan1993lattice,shan1994simulation,succi2001lattice}. We consider a binary immiscible mixture, where the two fluids have the same density and viscosity. We perform direct numerical simulations in a 3-dimensional and 3-periodic discretized computational domain of size $512\textsuperscript{3}$ to provide both qualitative morphological structures and quantitative information on droplet size during phase inversion. The emulsion is performed by mimicking the hydrodynamic stirring with a homogeneous isotropic injected forcing. 
The repulsive potential at the droplet interfaces is of Shan-Chen ~\cite{shan1993lattice,shan1994simulation} type while the disjoining potential is modeled as in~\cite{benzi2009mesoscopic,sbragaglia2012emergence} to avoid droplet coalescence at rest.
The increasing volume fraction of the dispersed droplet phase is achieved adiabatically adding the droplet component while consistently removing the contiguous fluid component. Details of the entire process of emulsification of dense emulsions via computer simulations are reported in a previous publication~\citep{girotto2022build}.

% Appendix
\section{SI Appendix}

\subsection*{A. The oil volume fraction}

\begin{figure*}[!htb]
	\centering
	\includegraphics[width = 0.65\textwidth]{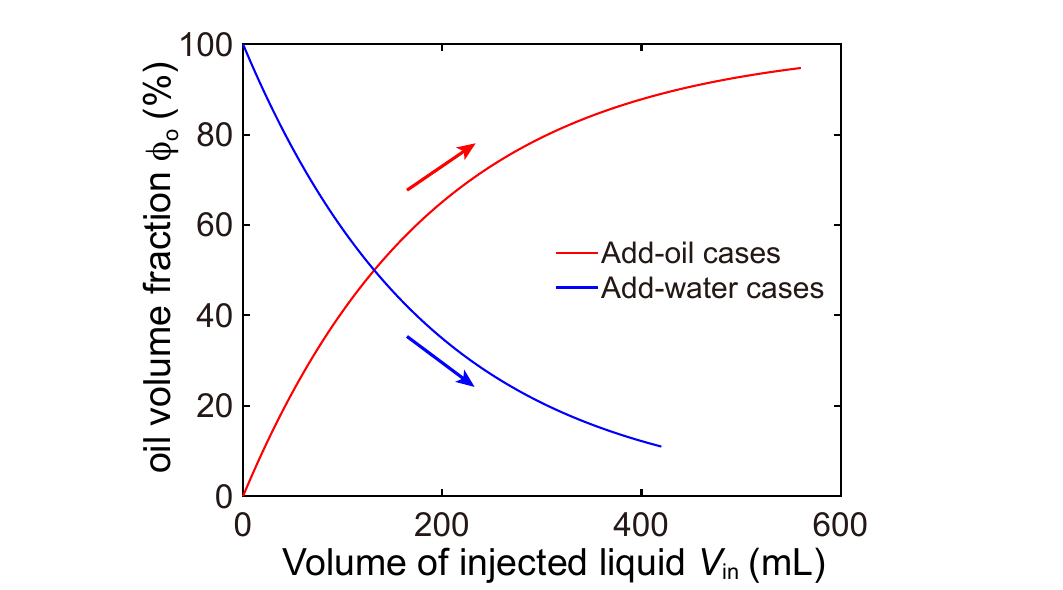}
	\caption{
	The oil volume fraction in the emulsion versus the volume of injected liquid for both the add-oil and add-water cases.
	}
	\label{fig1-SI}
\end{figure*}

In continuous experiments, we started from pure ethanol-water at $\phi_{o}=0\%$ (or pure oil at $\phi_{o}=100\%$), and dynamically injected oil (or ethanol-water) into the system to increase the volume fraction of the dispersed oil (or ethanol-water).
The inlet and outlet micropumps are set to work at the same constant volume flow rate $Q$, ensuring that the total liquid volume in the system $V_\text{tol}$ remains unchanged (see figure~\ref{fig1-SI} (a) in the main manuscript).
$V_\text{tol}$ and $Q$ are known, therefore the dispersed phase volume fraction as a function of the time (or injected liquid volume $V_\text{in}$) can be analytically derived as:
\begin{equation}
\phi_\text{dis}=1-\left(1-\phi_{i}\right) e^{-\frac{Q}{V_\text{tol}} t} = 1-\left(1-\phi_{i}\right) e^{-\frac{V_\text{in}}{V_\text{tol}}},
\label{equ1}
\end{equation}
where $\phi_{i}$ is the initial dispersed phase volume fraction before the injection and $t$ is the time from the start of injection.
As the experiments started from pure oil (or ethanol-water), we obtained $\phi_{i} = 0\%$ (or $\phi_{i} = 100\%$).
For all experiments discussed in the main manuscript, the input flow rate is $Q = \SI{10}{mL/min}$ and the total volume $V_\text{tol} = \SI{190}{mL}$.
The oil volume fraction as a function of the time is shown in figure~\ref{fig1-SI}, where the red line denotes the add-oil case while the blue line for the add-water case.
The final volume of injected liquid is high enough to make sure that the dispersed phase volume fraction is above the phase inversion point. Thus, the evolution of the emulsion across the phase inversion process can be measured.
For add-oil cases and add-water cases, the final volumes of injected liquid are set at $V_\text{in} = \SI{560}{mL}$ ($\phi_{o} = 95\%$) and at $V_\text{in} = \SI{420}{mL}$ ($\phi_{o} = 11\%$), respectively.

\subsection*{B. Torque measurements}

\begin{figure*}[!htb]
	\centering
	\includegraphics[width = 0.65\textwidth]{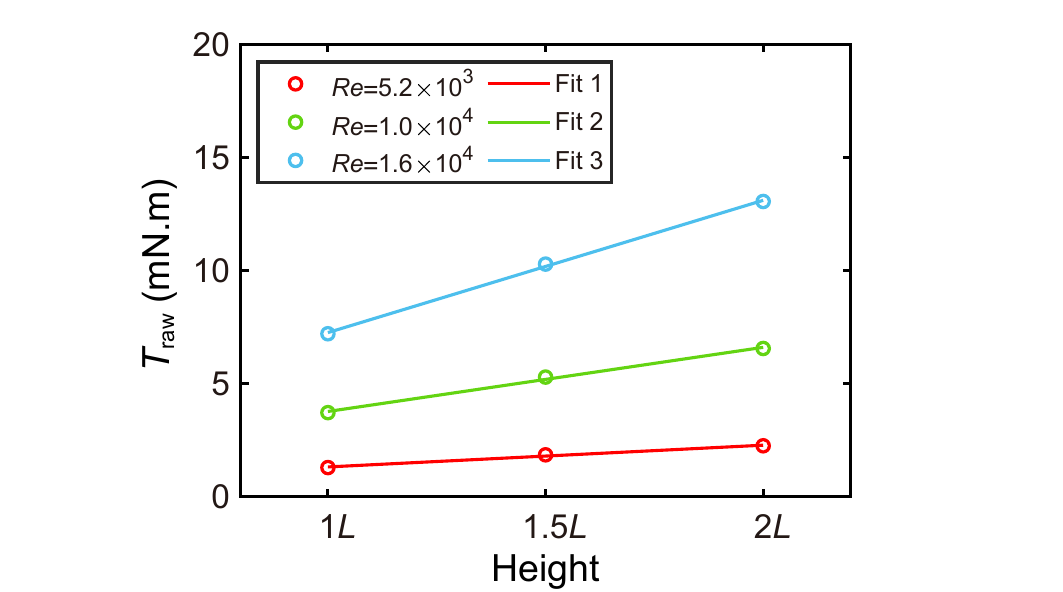}
	\caption{
	The total torque versus the height of the TC device. The torque contribution of the end effect can be determined as the longitudinal intercept of the fitting line. Here, the oil volume fraction is $\phi_{o} = 0\%$.
	}
	\label{fig2-SI}
\end{figure*}

In experiments, the torque, which is required to maintain the inner cylinder rotating at a constant angular velocity $\omega_{i}$, is directly measured by the torque sensor through the shaft connected to the rotating inner cylinder, and the accuracy is up to $0.1\rm~nN\cdot m$. 
The torque as a function of the time is recorded, which can be transferred into the torque versus the oil volume fraction.

Note that the total torque $T_\text{raw}$ contains two parts: (1) the torque $T$ due to the sidewall of the inner cylinder (Taylor-Couette flow); (2) the torque $T_\text{end}$ from the top and bottom end-plates (von K\'arm\'an flow). In this study, what we focus on is $T$, which can be determined using a linearization method~\citep{hu2017significant,greidanus2011drag}.
Here we show the details.
We performed torque measurements in three TC devices with different heights of, $L$, $2L$, and $3L$, while other parameters are the same. It has been found that the contribution of the cylindrical sidewall increases linearly with the height of the cylinder~\citep{greidanus2011drag,hu2017significant}, thus we can obtain the $T_\text{end}$ as the longitudinal intercept of the linear fit (see figure~\ref{fig2-SI}). The ratio of the torque caused by the TC flow to the total torque can be given as $\beta = 1 - T_\text{end}/T_\text{raw}$, which is determined by performing experiments for two cases of single-phase flow (i.e., $\phi_{o} = 0\%$ and $\phi_{o} = 100\%$). 
Then, $\beta$ obtained can be applied to the flow with internal dispersed phase (i.e., emulsions) provided that the emulsion is well-mixed. 
Consequently, the value of the torque $T$ can be calculated for various oil volume fractions.

To analyze the properties of the torque near the phase inversion, we decompose the torque $T$ as
\begin{equation}
T=\langle T\rangle+T^{\prime},
\end{equation}
where $\left<T\right>$ is the time-averaged component and $T^{\prime}$ the fluctuation component.

\subsection*{C. The effect of the injection flow rate}
\begin{figure*}[!htb]
	\centering
	\includegraphics[width = 0.65\textwidth]{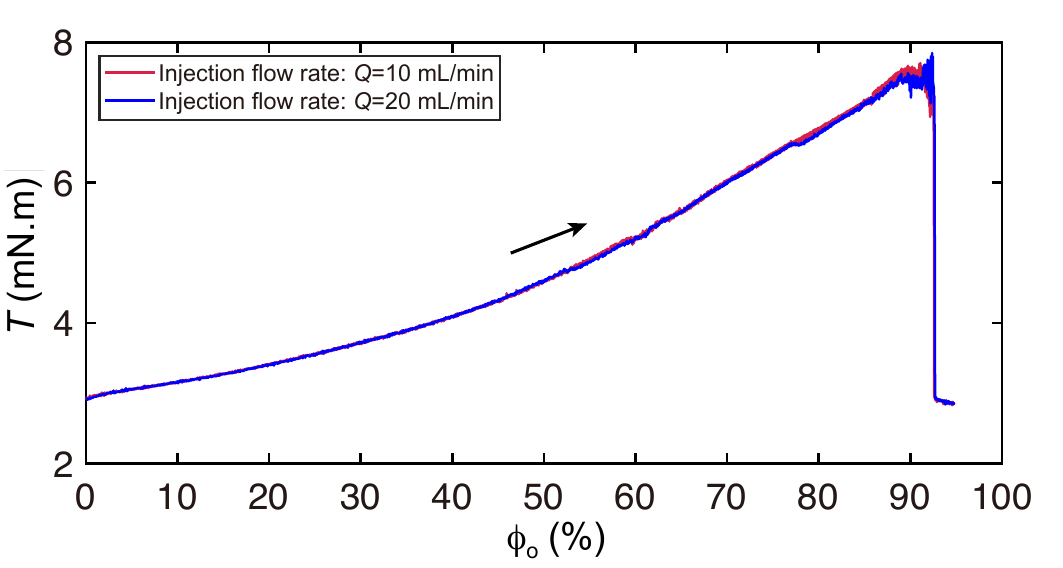}
	\caption{
	The torque $T$ versus the oil volume fraction $\phi_{o}$ for various injection flow rates. 
	}
	\label{fig3-SI}
\end{figure*}

We investigated the effect of the injection flow rate on the torque behavior at $Re = 1.0\times10^{4}$. Two different injection flow rates were used: $Q = \SI{10}{mL/min}$ and $Q = \SI{20}{mL/min}$.
As shown in figure~\ref{fig3-SI}, the global torque increases with the oil volume fraction until the phase inversion point, at which the torque undergoes a sudden drop.
The torque evolution for $Q = \SI{10}{mL/min}$ case and that for $Q = \SI{20}{mL/min}$ case almost collapses, indicating that the effect of the injection rate is minimal.
This result suggests that the $Q = \SI{10}{mL/min}$ experiments discussed in the main manuscript are in a quasi-static state.
In addition, the phase inversion points for different injection flow rates are almost the same.

\subsection*{D. Some additional snapshots for the phase inversion}
\begin{figure*}[!htb]
	\centering
	\includegraphics[width = 0.85\textwidth]{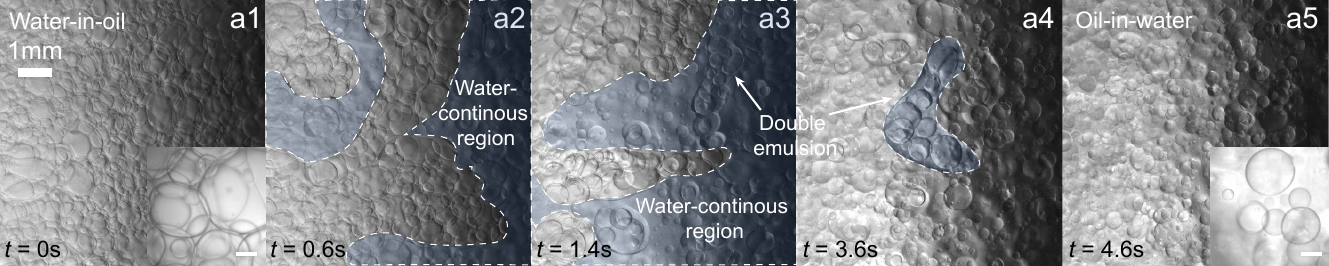}
	\caption{
	Some additional experimental snapshots for the phase inversion process of the add-water case (from W/O to O/W) with $Q = \SI{10}{mL/min}$ at $Re = 5.2\times10^{3}$.
	The light blue areas in (a2-a4) show the development of water-continuous regions. The scale bars in inset snapshots of both a1 and a5 denote $\SI{200}{\mu m}$.
	}
	\label{fig5-SI}
\end{figure*}
Here, we provide some additional snapshots to describe the phase inversion process in a relatively macroscopic view (high-speed camera Photron Fastcam AX200 mounted with macro lens Nikon 105~mm).
As shown in figure~\ref{fig5-SI}, the development of the water-continuous regions and the double emulsions are observed during the phase inversion.
We note that these unstable double emulsions only last for a short time during the phase inversion and they disappear quickly after the completion of the phase inversion.

\bibliographystyle{prsty_withtitle} 

%\bibliography{phase_inversion.bib}

\end{document}